\documentclass[11pt]{revtex4}
\usepackage{amssymb,epsf}
\usepackage{latexsym}
\begin{document}

\title{Thermodynamics of Charged Brans-Dicke AdS Black Holes}
\author{A. Sheykhi$^{1,2}$ \footnote{sheykhi@mail.uk.ac.ir} and  M. M. Yazdanpanah$^{1}$ \footnote{myazdan@mail.uk.ac.ir}}
\address{$^1$Department of Physics, Shahid Bahonar University, P.O. Box 76175, Kerman, Iran\\
         $^2$Research Institute for Astronomy and Astrophysics of Maragha (RIAAM), Maragha, Iran}
\begin{abstract}
It is well-known that in four dimensions, black hole solution of
the Brans-Dikce-Maxwell equations is just the Reissner-Nordstrom
solution with a constant scalar field. However, in $n\geq4$
dimensions, the solution is not yet the $(n+1)$-dimensional
Reissner-Nordstrom solution and the scalar field is not a constant
in general. In this paper, by applying a conformal transformation
to the dilaton gravity theory, we derive a class of black hole
solutions in $(n+1)$-dimensional $(n\geq 4)$ Brans-Dikce-Maxwell
theory in the background of anti-de Sitter universe. We obtain the
conserved and thermodynamic quantities through the use of the
Euclidean action method. We find a Smarr-type formula and perform
a stability analysis in the canonical ensemble. We find that the
solution is thermally stable for small $\alpha$, while for large
$\alpha$ the system has an unstable phase, where $\alpha $ is a
coupling constant between the scalar and matter field.
\end{abstract}
\maketitle
\section{Introduction\label{Intr}}
The pioneering study on scalar-tensor theories was done by Brans
and Dicke several decades ago who sought to incorporate Mach's
principle into gravity \cite{BD}. According to Brans-Dicke (BD)
theory the phenomenon of inertia arises from accelerations with
respect to the general mass distribution of the universe. This
theory can be regarded as an economic modification of general
relativity which accomodates both Mach's principle and Dirac's
large number hypothesis as new ingredients. There has been a
renewed interest in studying BD theory ever since it has been
disclosed that BD theory appears naturally in the low energy limit
of superstring theory. In string theory, gravity becomes
scalar-tensor in nature. The low-energy effective action of the
string theory leads to the Einstein gravity, coupled non-minimally
to a scalar dilaton field. Besides, recent observations show that
at the present epoch, our Universe expands with acceleration
instead of deceleration along the scheme of standard Friedmann
models and since general relativity could not describe such
Universe correctly, cosmologists have attended to alternative
theories of gravity such as BD theory. Due to highly nonlinear
character of BD theory, a desirable pre-requisite for studying
strong field situation is to have knowledge of exact explicit
solutions of the field equations. Since black holes are very
important both in classical and quantum gravity, many authors have
investigated various aspects of them in BD theory \cite{Sen}. It
turned out that the dynamic scalar field in the BD theory plays an
important role in the process of collapse and critical phenomenon.
The first four-dimensional black hole solutions of BD theory was
obtained by Brans in four classes \cite{Brans}. It has been shown
that among these four classes of the static spherically symmetric
solutions of the vacuum BD theory of gravity only two are really
independent, and only one of them is permitted for all values of
$\omega$. It has been proved that in four dimensions, the
stationary and vacuum BD solution is just the Kerr solution with
constant scalar field everywhere \cite{Hawking}. It has also been
shown that the charged black hole solution in four-dimensional
Brans-Dicke-Maxwell (BDM) theory is just the Reissner-Nordstrom
solution with a constant scalar field, however, in higher
dimensions, one obtains the black hole solutions with a nontrivial
scalar field \cite{Cai1}. This is because the stress energy tensor
of Maxwell field is not traceless in the higher dimensions and the
action of Maxwell field is not invariant under conformal
transformations. Accordingly, the Maxwell field can be regarded as
the source of the scalar field in the BD theory \cite{Cai1}. The
properties of charged black hole solutions in dilaton gravity
\cite{Cai4,Sheykhi0,Shey1,Shey2} and BD theory \cite{Shey3} have
been explored by many authors. However, these solutions
\cite{Cai4,Sheykhi0,Shey1,Shey2,Shey3} are neither asymptotically
flat nor (anti)-de Sitter [(A)dS]. Recently, the dilaton potential
leading to (A)dS-like solutions of dilaton gravity has been found
\cite{Gao1,Gao2,Gao3}. It was shown that the cosmological constant
is coupled to the dilaton in a very nontrivial way. Other studies
on black hole solutions in BD theory have been carried out in
\cite{Kim,Deh1,Gao}. In this paper, we would like to construct
black hole solutions of BDM theory in the background of (A)dS
spaces in all higher dimensions for an arbitrary value of coupling
constant and investigate their properties. We also want to perform
a stability analysis and investigate the effect of the scalar
field on the thermal stability of the solutions.

The outline of this paper is as follows: In section \ref{Conf}, we
present the basic equations and the conformal transformation
between the action of the dilaton gravity theory and the BD
theory. In section \ref{Top}, we construct black hole solutions in
BDM theory and investigate their properties. In section
\ref{Therm}, we study the thermodynamical properties of the
solutions and calculate the conserved and thermodynamic quantities
of BD black holes. We also investigate the effect of the scalar
field on the thermal stability of the solutions in this section.
The last section is devoted to summary and discussion.
\section{ Field equations and Conformal Transformations}\label{Conf}
The action of the $(n+1)$-dimensional Brans-Dicke-Maxwell theory
with one scalar field $\Phi$ and a self-interacting potential
$V(\Phi)$ can be written as
\begin{eqnarray}
I_{G}&=&-\frac{1}{16\pi}\int_{\mathcal{M}}
d^{n+1}x\sqrt{-g}\left(\Phi {R}-\frac{\omega}{\Phi}(\nabla\Phi)^2
-V(\Phi)-F_{\mu \nu}F^{\mu \nu}\right)\nonumber \\
&&-\frac{1}{8\pi}\int_{\partial{\mathcal{M}}}{d^{n}x\sqrt{-h}\Phi
K},\label{act1}
\end{eqnarray}
where ${R}$ is the scalar curvature, $V(\Phi )$ is a potential for
the scalar field $\Phi $, $F_{\mu \nu }=\partial _{\mu }A_{\nu
}-\partial _{\nu }A_{\mu }$ is the electromagnetic field tensor,
and $A_{\mu }$ is the electromagnetic potential. The factor
$\omega$ is the coupling constant. The last term in Eq.
(\ref{act1}) is the Gibbons-Hawking boundary term which is
chosen such that the variational principle is well-defined. The manifold $%
\mathcal{M}$ has metric $g_{\mu \nu }$ and covariant derivative
$\nabla _{\mu }$. $K$ is the trace of the extrinsic curvature $K
^{ab}$ of the boundary $\partial \mathcal{M}$ of the manifold
$\mathcal{M}$, with induced metric $h_{ab}$. The equations of
motion can be obtained by varying the action (\ref{act1}) with
respect to the gravitational field $g_{\mu \nu }$, the scalar
field $\Phi $ and the gauge field $A_{\mu }$ which yields the
following field equations
\begin{eqnarray}
&&G_{\mu
\nu}=\frac{\omega}{\Phi^2}\left(\nabla_{\mu}\Phi\nabla_{\nu}\Phi-\frac{1}{2}g_{\mu
\nu}(\nabla\Phi)^2\right)
-\frac{V(\Phi)}{2\Phi}g_{\mu \nu}+\frac{1}{\Phi}\left(\nabla_{\mu}\nabla_{\nu}\Phi-g_{\mu \nu}\nabla^2\Phi\right)\nonumber \\
&&+\frac{2}{\Phi}\left(F_{\mu \lambda}F_{ \nu}^{\
\lambda}-\frac{1}{4}F_{\rho \sigma}F^{\rho
\sigma}g_{\mu \nu}\right), \label{Eq1}\\
&&\nabla^2\Phi=-\frac{n-3}{2(n-1)\omega+2n}F^2+\frac{1}{2(n-1)\omega+2n}\left((n-1)\Phi\frac{dV(\Phi)}{d\Phi}
-(n+1)V(\Phi)\right),\label{Eq2} \\
&&\nabla_{\mu}F^{\mu \nu}=0, \label{Eq3}
\end{eqnarray}
where $G_{\mu \nu}$ and $\nabla$ are, respectively, the Einstein
tensor and covariant differentiation in the spacetime metric
$g_{\mu \nu}$. It is apparent that the right hand side of Eq.
(\ref{Eq1}) includes the second derivatives of the scalar field,
so it is hard to solve the field equations (\ref{Eq1})-(\ref{Eq3})
directly. We can remove this difficulty by a conformal
transformation. Indeed, the BDM theory (\ref{act1}) can be
transformed into the Einstein-Maxwell theory with a minimally
coupled scalar dilaton field, $\bar{\Phi}$, via the conformal
transformation \cite{Cai1}
\begin{eqnarray} \label{conf}
&&\bar{g}_{\mu \nu}=\Omega^ {-2}g_{\mu \nu},
\end{eqnarray}
with
\begin{eqnarray}
&&\Omega^ {-2}=\Phi^{\frac{2}{n-1}},
\end{eqnarray}
and
\begin{equation}
\alpha=\frac{n-3}{\sqrt{4(n-1)\omega+4n}}, \  \   \
\bar{\Phi}=\frac{n-3}{4\alpha}\ln \Phi. \label{6}
\end{equation}
Using this conformal transformation, the action (\ref{act1})
transforms to
\begin{eqnarray}
\bar{I}_{G}&=&-\frac{1}{16\pi}\int_{\mathcal{M}}
d^{n+1}x\sqrt{-\bar{g}}\left({\bar{R}}-\frac{4}{n-1}(\bar{\nabla}\
\bar{\Phi})^2-\bar{V}(\bar{\Phi})-e^{-\frac{4\alpha\bar{\Phi}}{n-1}}\bar{F}_{\mu
\nu}\bar{F}^{\mu \nu}\right)\nonumber \\
&&-\frac{1}{8\pi}\int_{\partial{\mathcal{M}}}{d^{n}x\sqrt{-\bar{h}}\bar{
K}}, \label{act2}
\end{eqnarray}
where ${\bar{R}}$ and $\bar{\nabla}$ are, respectively, the Ricci
scalar and covariant differentiation in the spacetime metric
$\bar{g}_{\mu \nu}$, and $\bar{V}(\bar{\Phi})$ is
\begin{equation}
\bar{V}(\bar{\Phi})=\Phi^{-\frac{n+1}{n-1}}V(\Phi).\label{8}
\end{equation}
This action is just the action of the $(n+1)$-dimensional
Einstein-Maxwell-dilaton gravity, where $\bar{\Phi}$ is the
dilaton field and $\bar{V}(\bar{\Phi})$ is a potential for
$\bar{\Phi}$. $\alpha $ is an arbitrary constant governing the
strength of the coupling between the dilaton and the Maxwell
field. Varying action (\ref{act2}), we obtain the equations of
motion
\begin{eqnarray}
&&\bar{{R}}_{\mu
\nu}=\frac{4}{n-1}\left(\bar{\nabla}_{\mu}\bar{\Phi}\bar{\nabla}
_{\nu}\bar{\Phi}+\frac{1}{4}\bar{V}(\bar{\Phi})\bar{g}_{\mu
\nu}\right)+ 2e^{\frac{-4\alpha\bar{\Phi}}{n-1}}\left(\bar{F}_{\mu
\lambda}\bar{F}_{\nu}^{ \ \lambda} -\frac{1}{2(n-1)}\bar{F}_{\rho
\sigma}\bar{F}^{\rho \sigma}\bar{g}_{\mu \nu}\right) \label{Eqd1}\\
&&
\bar{\nabla}^2\bar{\Phi}=\frac{n-1}{8}\frac{\partial\bar{V}}{\partial\bar{\Phi}}
-\frac{\alpha}{2}e^{\frac{-4\alpha\bar{\Phi}}{n-1}}\bar{F}_{\rho
\sigma}\bar{F}^{\rho \sigma},\label{Eqd2}\\
&&
\bar{\nabla}_{\mu}\left(e^{\frac{-4\alpha\bar{\Phi}}{n-1}}\bar{F}^{\mu
\nu}\right)=0. \label{Eqd3}
\end{eqnarray}
Comparing Eqs. (\ref{Eq1})-(\ref{Eq3}) with Eqs.
(\ref{Eqd1})-(\ref{Eqd3}), we find that if $\left(\bar{g}_{\mu
\nu},\bar{F}_{\mu \nu},\bar{\Phi}\right)$ is the solution of Eqs.
(\ref{Eqd1})-(\ref{Eqd3}) with potential $\bar{V}(\bar{\Phi})$,
then
\begin{equation}\label{conform2}
\left[{g}_{\mu \nu},{F}_{\mu
\nu},{\Phi}\right]=\left[\exp\left({\frac{-8\alpha
\bar{\Phi}}{(n-1)(n-3)}}\right)\bar{g}_{\mu \nu},\bar{F}_{\mu
\nu},\exp\left({\frac{4\alpha \bar{\Phi}}{n-3}}\right)\right],
\end{equation}
is the solution of Eqs. (\ref{Eq1})-(\ref{Eq3}) with potential
$V(\Phi)$.
\section{Brans-Dicke black holes in AdS spaces\label{Top}}
Asymptotically (A)dS-like solutions of the dilaton field equations
(\ref{Eqd1})-(\ref{Eqd3}) have been constructed in
\cite{Gao1,Gao2,Gao3,Ghosh,Sheykhi1,Sheykhi2}. Here we would like
to obtain  black hole solutions of the Brans-Dicke field equations
(\ref{Eq1})-(\ref{Eq3}) in the background of AdS universe. Our
strategy for constructing these solutions is applying the
conformal transformation (\ref{conform2}) to black hole solutions
of Eqs. (\ref{Eqd1})-(\ref{Eqd3}) in the dilaton gravity theory.
The dilaton potential leading to (A)dS-like solutions of dilaton
gravity has been found recently in \cite{Gao2}. For an arbitrary
value of $\alpha $ in (A)dS spaces the form of the dilaton
potential $\bar{V}(\bar{\Phi})$ in $(n+1)$-dimensions is chosen as
\begin{eqnarray}\label{V1}
\bar{V}(\bar{\Phi})&=&\frac{\Lambda(n-1)}{3(n-2+\alpha^2)^{2}}\Bigg{\{}-\alpha^2\left[(n+1)^{2}-(n+1)\alpha^{2}-6(n+1)+\alpha^{2}+9\right]
e^{\frac{-4(n-2)\bar{\Phi}}{(n-1)\alpha}} \nonumber
\\
&& +(n-2)^{2}(n-\alpha^{2})
e^{\frac{4\alpha\bar{\Phi}}{n-1}}+4\alpha^{2}(n-1)(n-2)
e^{\frac{-2\bar{\Phi}(n-2-\alpha^{2})}{(n-1)\alpha}}\Bigg {\}}.
\end{eqnarray}
Here $\Lambda $ is the cosmological constant. It is clear the
cosmological constant is coupled to the dilaton in a very
nontrivial way. This type of dilaton potential can be obtained
when a higher dimensional theory is compactified to four
dimensions, including various supergravity models \cite{Gid}. In
the absence of the dilaton field the action (\ref{act2}) reduces
to the action of Einstein-Maxwell gravity with cosmological
constant. Asymptotically AdS black hole solutions of the field
equations (\ref{Eqd1})-(\ref{Eqd3}) have been obtained in
\cite{Gao2} and we review it briefly here. Assuming the
$(n+1)$-dimensional metric has the following form
\begin{equation}
d\bar{s}^2=-f(r)dt^2+\frac{dr^2}{g(r)}+r^2{R^2(r)}d\Omega^2_{n-1},
\label{met1}
\end{equation}
where $d\Omega^2_{n-1}$ denotes the metric of an unit
$(n-1)$-sphere and $f(r)$, $g(r)$ and $R(r)$ are functions of $r$
which should be determined. First of all, the Maxwell equations
(\ref{Eqd3}) can be integrated immediately, where all the
components of $\bar{F}_{\mu\nu}$ are zero except $\bar{F}_{tr}$:
\begin{equation}\label{Ftr}
\bar{F}_{tr}=\sqrt{\frac{f(r)}{g(r)}}\frac{q e^{\frac{4\alpha
\bar{\Phi}}{n-1}}}{\left( rR\right) ^{n-1}} ,
\end{equation}
where $q$, an integration constant, is the charge parameter of the
black hole. According to the Gauss theorem, the electric charge is
\begin{equation}\label{Q}
 Q = \frac{1}{4\pi} \int_{r\rightarrow \infty}
\bar{F}_{tr} \sqrt{-\bar{g}} d^{n-1}x=\frac{\Omega _{n-1}}{4\pi}q.
\end{equation}
where $\Omega _{n-1}$ is the volume of the unit $(n-1)$-sphere.
Notice that $Q$ is invariant under the conformal transformation
(\ref{conform2}). Using metric (\ref{met1}) and the Maxwell field
(\ref{Ftr}), one can show that the system of equations
(\ref{Eqd1})-(\ref{Eqd2}) have solutions of the form \cite{Gao2}
\begin{eqnarray}\label{f}
f(r)&=&\left[1-\left(\frac{r_{+}}{r}\right)^{n-2}\right]\left[1-\left(\frac{r_{-}}{r}\right)^{n-2}\right]^{1-\gamma\left(n-2\right)}-\frac{1}{3}\Lambda
r^2\left[1-\left(\frac{r_{-}}{r}\right)^{n-2}\right]^{\gamma} ,\\
g(r)&=&\Bigg{\{}\left[1-\left(\frac{r_{+}}{r}\right)^{n-2}\right]
\left[1-\left(\frac{r_{-}}{r}\right)^{n-2}\right]^{1-\gamma\left(n-2\right)}-\frac{1}{3}\Lambda
r^2\left[1-\left(\frac{r_{-}}{r}\right)^{n-2}\right]^{\gamma}\Bigg
{\}}\nonumber
\\&& \times
\left[1-\left(\frac{r_{-}}{r}\right)^{n-2}\right]^{\gamma(n-3)},\label{g}\\
\bar{\Phi}(r)&=&\frac{n-1}{4}\sqrt{\gamma(2+2\gamma-n\gamma)}\ln\left[1-\left(\frac{r_{-}}{r}\right)^{n-2}\right],\label{Phi}\\
R(r)&=&\left[1-\left(\frac{r_{-}}{r}\right)^{n-2}\right]^{\gamma/2},\label{R}
\end{eqnarray}
Here $r_+$ and $r_{-}$ are, respectively, the event horizon and
Cauchy horizon of the black hole, and the constant $\gamma$ is
\begin{equation}\label{gamma}
\gamma=\frac{2\alpha^{2}}{(n-2)(n-2+\alpha^{2})}.
\end{equation}
The charge parameter $q$ is related to $r_+$ and $r_{-}$ by
\begin{equation}\label{q}
q^{2}=\frac{(n-1)(n-2)^{2}}{2(n-2+\alpha^{2})}r_{+}^{n-2}r_{-}^{n-2}.
\end{equation}
The quasilocal mass  the dilaton AdS black hole can be calculated
through the use of the subtraction method of Brown and York
\cite{BY}. Such a procedure causes the resulting physical
quantities to depend on the choice of reference background.
According to this formalism if we write the metric of spherically
symmetric spacetime in the form
\begin{equation}\label{metric2}
ds^2=-W(r)dt^2 + {dr^2\over V(r)}+ r^2d\Omega^2_{n-1},
\end{equation}
and the matter action contains no derivatives of the metric, then
the quasilocal mass is given by \cite{CHM}
\begin{equation}\label{QLM}
{\cal M} = \frac{n-1}{2}r^{n-2}{W^{1/2}(r)}\left( {V_{0}^{1/2}(r)}
- {V^{1/2}(r)}\right).
\end{equation}
Here $V_{0}(r)$ is an arbitrary function which determines the zero
of the energy for a background spacetime and $r$ is the radius of
the spacelike hypersurface boundary. When the spacetime is
asymptotically (A)dS, the Arnowitt-Deser-Misner (ADM) mass $M$ is
the ${\cal M}$ determined in (\ref{QLM}) in the limit
$r\rightarrow\infty$. If no cosmological horizon is present, the
large $r$ limit of (\ref{QLM}), is used to determine the mass.  If
a cosmological horizon is present one can not take the large $r$
limit to identify the quasilocal mass. However, one can still
identify the small mass parameter in the solution \cite{BY}. For
the solution under consideration, there is no cosmological horizon
and if we transform the metric (\ref{met1}) in the form
(\ref{metric2}) by using the transformation
\begin{equation}\label{tran1}
r^2 R^2(r)\rightarrow r^2,
\end{equation}
then we obtain the mass of the dilaton black hole as
\begin{equation}\label{mass}
\bar{M}=\frac{\Omega
_{n-1}}{16\pi}(n-1)\left[r^{n-2}_{+}+\frac{n-2-\alpha^2}{n-2+\alpha^2}r^{n-2}_{-}\right].
\end{equation}
In the absence of the dilaton field ($\alpha=0$) this expression
for the mass reduces to the mass of the $(n+1)$-dimensional
Reissner-Nordstrom-AdS black holes. It is worth noting that our
result for ADM mass coincides with ones found in \cite{Fang} for
asymptotically flat dilaton black holes. Next we calculate the
entropy of the dilatonic black hole. Black hole entropy typically
satisfies the so called area law of the entropy, which states that
the entropy of the black hole is a quarter of the event horizon
area \cite{Beck}. This near-universal law applies to almost all
kinds of black holes, including dilaton black holes, in Einstein
gravity \cite{hunt}. It is easy to show that the entropy of the
black hole is
\begin{equation}\label{Entropy1}
\bar{S}=\frac{\Omega_{n-1}r_{+}^{n-1}}{4}\left[1-\left(\frac{r_{-}}{r_{+}}\right)^{n-2}\right]^{\gamma
(n-1)/2}.
\end{equation}
The Hawking temperature of the  black hole on the outer horizon
$r_+$, in dilaton gravity, can be calculated using the relation
\begin{equation}
\bar{T}_{+}= \left(\frac{f^{\text{ }^{\prime
}}}{4\pi\sqrt{f/g}}\right)_{r=r_{+}},
\end{equation}
where a prime denotes derivative with respect to $r$. It is a
matter of calculation to show  that
\begin{eqnarray}\label{Tem}
\bar{T}_{+}&=&\frac{(n-2)}{4\pi r_{+}} \left[ 1- \left({\frac
{r_{-}}{r_{+}}} \right) ^{n-2} \right]^{1-\gamma(n-1)/2}.
\end{eqnarray}
It is apparent that the metric corresponding to
(\ref{f})-(\ref{R}) is asymptotically (A)dS. Using the conformal
transformation (\ref{conform2}), the $(n+1)$-dimensional black
hole solutions of BDM theory in the background of AdS universe can
be obtained as
\begin{equation}
ds^2=-U(r)dt^2+\frac{dr^2}{V(r)}+r^2{H^2(r)}d\Omega^2_{n-1},
\label{met2}
\end{equation}
where $U(r)$, $V(r)$, $H(r)$ and $\Phi(r)$ are
\begin{eqnarray}
&&U(r)=\left[1-\left(\frac
{r_{-}}{r}\right)^{n-2}\right]^{\frac{-2(n-2)\gamma}{n-3}} f(r) ,
 \label{U1}\\
&&V(r)=\left[1-\left(\frac
{r_{-}}{r}\right)^{n-2}\right]^{\frac{2(n-2)\gamma}{n-3}} g(r) ,
\label{V1}\\
&& H(r)=\left[1-\left(\frac
{r_{-}}{r}\right)^{n-2}\right]^{\frac{-(n-1)\gamma}{2(n-3)}},
\label{H1}\\
&&\Phi(r)=\left[1-\left(\frac
{r_{-}}{r}\right)^{n-2}\right]^{\frac{(n-1)(n-2)\gamma}{(n-3)}}.
\label{Phi1}
\end{eqnarray}
Applying the conformal transformation (\ref{conform2}), the
electromagnetic field in BDM theory can be written as
\begin{eqnarray}
F_{tr}=\bar{F}_{tr}=\frac{q}{r^{n-1}}, \label{Ftr2}
\end{eqnarray}
while the scalar potential in  BDM theory becomes
\begin{eqnarray}
V(\Phi)&=&\frac{\Lambda(n-1)}{3(n-2+\alpha^2)^{2}}\Bigg{\{}4\,{\alpha}^{2}
\left( n-1 \right) \left( n-2 \right) {\Phi}^{{\frac {(3n-1)
{\alpha}^{2}-(n-2)(n-3)}{2{\alpha}^{2} \left( n-1 \right) }}}
\nonumber
\\
&& +
 \left( n-2 \right) ^{2} \left( n-{\alpha}^{2} \right){\Phi}^{2} +{\alpha}^{ 2} \left[ n{\alpha}^{2}- \left( n-2 \right) ^{2}
\right]{\Phi}^{{\frac {( n+1) {\alpha}^{2}-
(n-2)(n-3)}{{\alpha}^{2} \left( n-1 \right) }}}\Bigg
{\}}.\label{V2}
\end{eqnarray}
As one can see from Eq. (\ref{Ftr2}), in the background of (A)dS
universe, the scalar field in BD theory does not exert any direct
influence on the matter field $F_{tr}$, however,  the scalar field
modifies the geometry of the spacetime as it participate in the
field equations. This is in contrast to the solutions presented in
\cite{Deh1,Shey3}. The solutions in \cite{Deh1,Shey3} are neither
asymptotically flat nor (A)dS  and the gauge field crucially
depends on the scalar BD field. It is worth noting that the scalar
field $\Phi(r)$ and the electromagnetic field $F_{tr}$ become zero
as $r\rightarrow\infty$. It is also notable to mention that these
solutions are valid for all values of $\omega$. When
$\omega\rightarrow\infty$ ($\alpha=0=\gamma$), these solutions
reduce to
\begin{eqnarray}
U(r)=V(r)=\left[1-\left(\frac{r_{+}}{r}\right)^{n-2}\right]\left[1-\left(\frac{r_{-}}{r}\right)^{n-2}\right]-\frac{1}{3}\Lambda
r^2,
\end{eqnarray}
which describes an $(n+1)$-dimensional asymptotically (A)dS
Reissner-Nordstrom black hole.
\section{Thermodynamics of BD black holes \label{Therm}}
We now turn to the investigation of the thermodynamics of charged
BD black holes we have just found. The Hawking temperature of BD
black holes on the outer horizon $r_+$ can be calculated using the
relation
\begin{equation}
T_{+}=\frac{\kappa}{2\pi}= \left(\frac{U^{\text{ }^{\prime
}}}{4\pi\sqrt{U/V}}\right)_{r=r_{+}},
\end{equation}
where $\kappa$ is the surface gravity. We obtain
\begin{eqnarray}\label{TemBD}
T_{+}&=&\frac{(n-2)}{4\pi r_{+}} \left[ 1- \left({\frac
{r_{-}}{r_{+}}} \right) ^{n-2} \right]^{1-\gamma(n-1)/2}.
\end{eqnarray}
If we compare Eq. (\ref{TemBD}) with the temperature obtained in
the dilaton gravity theory in Eq. (\ref{Tem}), we find that the
temperature is invariant under the conformal transformation
(\ref{conform2}). This is due to the fact that the conformal
parameter $\Omega^2$ is regular at the horizon. Therefore, the
Hawking temperature is an invariant quantity under conformal
transformations only if the transformations are regular at event
horizon.

The ADM mass $M$, the entropy $S$ and the electric potential $U$
of the BD black hole can be calculated through the use of the
Euclidean action method \cite{BCR,CaiSu}. In this approach, first
the electric potential and the temperature are fixed on a boundary
with a fixed radius $r_{+}$. The Euclidean action has two parts;
bulk and surface. The first step to make the Euclidean action is
to substitute $t$ with $i\tau$. This makes the metric positive
definite:
\begin{equation}
ds^2=U(r)d\tau^2+\frac{1}{V(r)}dr^2+r^2H^2(r)d\Omega^2_{n-1}.
\label{Eumetr}
\end{equation}
There is a conical singularity at the horizon $r=r_{+}$ in the
Euclidean metric \cite{CaiSu}. To eliminate it, the Euclidian time
$\tau$ is made periodic with period $\beta$, where $\beta$ is the
inverse of Hawking temperature. Now we obtain the Euclidean action
of $(n+1)$-dimensional Brans-Dicke-Maxwell theory. The Euclidean
action can be calculated analytically and continuously changing of
action (\ref{act1}) to Euclidean time $\tau$, i.e.,
\begin{equation}
I_{GE}=-\frac{1}{16\pi}\int_{\mathcal{M}}
d^{n+1}x\sqrt{g}\left(\Phi {R}-\frac{\omega}{\Phi}(\nabla\Phi)^2
-V(\Phi)-F_{\mu \nu}F^{\mu
\nu}\right)-\frac{1}{8\pi}\int_{\partial{\mathcal{M}}}{d^{n}x\sqrt{h}\Phi(K-K_{0})},\label{act1e}
\end{equation}
where $K_{0}$  is the trace of the extrinsic curvature of the
vacuum metric background (here it is the $(n+1)$-dimensional AdS
spacetime). This term must be added so that it can normalize the
Euclidean action to zero in this spacetime \cite{Brown1}. Using
metric (\ref{Eumetr}), we find
\begin{eqnarray}
R&=&-g^{-1/2}(g^{1/2}U^{\prime}V/U)^{\prime}-2G^{0}_{0},\label{RE}\\
K&=&-\frac{\sqrt {V} \left[rH U^{\prime}+2(n-1)\left(
UH+rUH^{\prime}\right)\right] }{2rHU}, \label{K}
\end{eqnarray} where $G^{0}_{0}$ is the (00)
component of the Einstein tensor, and again the prime denotes
derivative with respect to $r$. Inserting $U(r)$, $V(r)$ and
$H(r)$ from (\ref{U1})-(\ref{H1}) with $r_{+}=0=r_{-}$ in Eq.
(\ref{K}) we obtain the extrinsic curvature for the metric
background
\begin{eqnarray}\label{K0}
K_{0}&=&\frac
{1-n\left(1-\Lambda{r}^{2}/3\right)}{r\sqrt{1-\Lambda{r}^{2}/3}}.
\end{eqnarray}
For $\Lambda=0$, we get $K_{0}=-(n-1)/{r}$ which is the trace of
the extrinsic curvature of the metric background for
asymptotically flat spacetimes \cite{CaiSu}. Substituting Eqs.
(\ref{RE})-(\ref{K0}) in action (\ref{act1e}) and using Eqs.
(\ref{U1})-(\ref{V2}), after a long calculation, we obtain the
Euclidean action as
\begin{eqnarray}
I_{GE}&=&\beta\frac{\Omega
_{n-1}}{16\pi}(n-1)\left[r^{n-2}_{+}+\frac{n-2-\alpha^2}{n-2+\alpha^2}r^{n-2}_{-}\right]
\nonumber
\\
&&-\frac{\Omega_{n-1}r_{+}^{n-1}}{4}\left[1-\left(\frac{r_{-}}{r_{+}}\right)^{n-2}\right]^{\gamma
(n-1)/2}-\beta\frac{\Omega_{n-1}q^2}{4\pi(n-2) r_{+}^{n-2}},
\label{IE}
\end{eqnarray}
According to Refs. \cite{Brown1,Brown2,Brown4}, the
thermodynamical potential can be given by $I_{GE}$, we get
\begin{equation}
I_{GE}=\beta M-S-\beta UQ, \label{GD}
\end{equation}
where $M$ is the ADM mass, $S$ and $U$ are the entropy and the
electric potential, respectively. Comparing Eq. (\ref{IE}) with
Eq. (\ref{GD}), we find
\begin{equation}
{M}=\frac{\Omega
_{n-1}}{16\pi}(n-1)\left[r^{n-2}_{+}+\frac{n-2-\alpha^2}{n-2+\alpha^2}r^{n-2}_{-}\right],
\label{Mass}
\end{equation}
\begin{equation}
S=\frac{\Omega_{n-1}r_{+}^{n-1}}{4}\left[1-\left(\frac{r_{-}}{r_{+}}\right)^{n-2}\right]^{\gamma
(n-1)/2},\label{entropy}
\end{equation}
\begin{equation}
U=\frac{q}{(n-2)r_{+}^{n-2}}.\label{elect}
\end{equation}
Comparing the conserved and thermodynamic quantities calculated in
this section with those obtained in the previous section, we find
that they are invariant under the conformal transformation
(\ref{conform2}). This is because the Euclidean action is
invariant under the conformal transformation (up to a surface term
associated with the scalar field). It is worth emphasizing that in
BD theory, where we have the additional gravitational scalar
degree of freedom, the entropy of the black hole does not follow
the area law \cite{kang}. This is due to the fact that the black
hole entropy comes from the boundary term in the Euclidean action
formalism. Nevertheless, the entropy remains unchanged under the
conformal transformations. The advantage of the Euclidean action
method is that, in principle, we can find all the thermodynamic
quantities because in this method the characteristic thermodynamic
function, i.e., the thermodynamical potential, is found.

\begin{figure}[tbp]
\epsfxsize=7cm \centerline{\epsffile{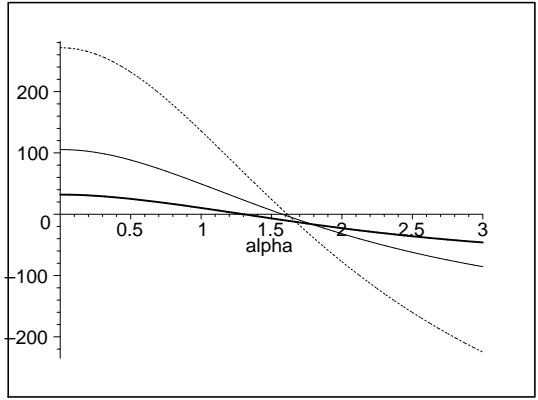}} \caption{$(\partial
^{2}M/\partial S^{2})_{Q}$ versus $\protect\alpha $ for
$r_{-}=0.2$ and $r_{+}=0.3$., $n=4$ (bold line), $n=5$ (continuous
line), and $n=6$ (dashed line).} \label{Fig}
\end{figure}

Then, we consider the first law of thermodynamics for the black
hole. In order to check the first law, we obtain the mass $M$ as a
function of extensive quantities $S$ and $Q$. Using the expression
for the charge, the mass and the entropy given in Eqs. (\ref{Q}),
(\ref{Mass}) and (\ref{entropy}), we can obtain a Smarr-type
formula as
\begin{equation}
M(S,Q)=\frac{\Omega
_{n-1}}{16\pi}(n-1)\left[Z+\frac{32Q^2(n-2-\alpha^2)\pi^2}{Z(n-1)(n-2)^2}\right],
\label{Msmarr}
\end{equation}
where $Z=r_{+}^{n-2}$ is the positive real root of the following
equation:
\begin{eqnarray}
Z^{\frac{n-1}{n-2}}\left[1-\frac{32Q^2(n-2+\alpha^2)\pi^2}{Z^{2}(n-1)(n-2)^2}\right]^{\gamma(n-1)/2}-4S=0.
\label{Zeq}
\end{eqnarray}
One may then regard the parameters $S$ and $Q$ as a complete set
of extensive parameters for the mass $M(S,Q)$ and define the
intensive parameters conjugate to $S$ and $Q$. These quantities
are the temperature and the electric potential
\begin{eqnarray}\label{inte1}
T&=&\left( \frac{\partial {M}}{\partial {S}}\right)_{Q}=
\left(\frac{\partial {M}}{\partial {Z}}\right)_{Q}\left( \frac{\partial {Z}}{\partial {S}}\right)_{Q},\\
U&=&\left(\frac{\partial {M}}{\partial {Q}}\right)_{S}=\left(
\frac{\partial {M}}{\partial {Q}}\right)_{Z}+\left( \frac{\partial
{M}}{\partial {Z}}\right)_{Q}\left( \frac{\partial {Z}}{\partial
{Q}}\right)_{S}.\label{inte2}
\end{eqnarray}
Straightforward calculations show that the intensive quantities
calculated by Eqs. (\ref{inte1}) and  (\ref{inte2}) coincide with
Eqs. (\ref{TemBD}) and (\ref{elect}). Thus, these thermodynamic
quantities satisfy the first law of black hole thermodynamics,
\begin{equation}
dM = TdS+Ud{Q}.
\end{equation}
Finally, we study the thermal stability of the solutions in the
canonical ensemble. The stability of a thermodynamic system with
respect to small variations of the thermodynamic coordinates is
usually performed by analyzing the behavior of the entropy $
S(M,Q)$ around the equilibrium. The local stability in any
ensemble requires that $S(M,Q)$ be a convex function of the
extensive variables or its Legendre transformation must be a
concave function of the intensive variables. The stability can
also be studied by the behavior of the energy $M(S,Q)$ which
should be a convex function of its extensive variable. Thus, the
local stability can in principle be carried out by finding the
determinant of the Hessian matrix of $M(S,Q)$ with respect to its
extensive variables $X_{i}$,
$\mathbf{H}_{X_{i}X_{j}}^{M}=[\partial ^{2}M/\partial
X_{i}\partial X_{j}]$ \cite{Cal2,Gub}. In our case the mass $M$ is
a function of entropy and charge. The number of thermodynamic
variables depends on the ensemble that is used. In the canonical
ensemble, the charge is a fixed parameter and therefore the
positivity of the $(\partial ^{2}M/\partial S^{2})_{Q}$ is
sufficient to ensure local stability. We have shown in figure
\ref{Fig} the behavior of $(\partial ^{2}M/\partial S^{2})_{Q}$
versus $\alpha$ in various dimensions. This figure shows that for
fixed value of the other parameters, the solution is thermally
stable for small value of $\alpha$ in any dimension, while it has
an unstable phase for large values of $\alpha$. This shows that
the scalar field makes the solution unstable.

\section{Summary and discussion \label{sum}}
The construction and analysis of the black hole solutions in
anti-de Sitter (AdS) spaces is a subject of much recent interest.
This interest is motivated by the correspondence between the
gravitating fields in an AdS spacetime and conformal field theory
on the boundary of the AdS spacetime. In this paper, with an
appropriate combination of three Liouville-type dilaton potentials
and applying a conformal transformation to the dilatonic black
hole solutions, we construct a class of $(n+1)$-dimensional
$(n\geq4)$  black hole solutions in BDM theory for arbitrary
values of the coupling constant $\omega$. These solutions are
asymptotically anti-de Sitter. We found the scalar potential
leading to AdS-like solutions in BDM theory. The cosmological
constant couples to the scalar field in a very nontrivial way, and
the scalar potential has a complicated form (see Eq. \ref{V2}).
This scalar potential plays a crucial role in the existence of
these black holes, as the negative cosmological constant does in
the Einstein-Maxwell theory. We found that the scalar field in BD
theory does not exert any direct influence on the gauge field
$F_{tr}$, however, the scalar field modifies the geometry of the
spacetime as it participates in the field equations. We obtained
the conserved and thermodynamic quantities through the use of the
Euclidean action method, and verified that the conserved and
thermodynamic quantities of the solutions satisfy the first law of
black hole thermodynamics. We found that the conserved and
thermodynamic quantities are invariant under the conformal
transformation. We also analyzed the thermal stability of the
solutions in the canonical ensemble by finding a Smarr-type
formula and considering $ (\partial ^{2}M/\partial S^{2})_{Q}$ for
the charged BD black hole solutions in $(n+1)$ dimensions. We
found that there is no Hawking-Page phase transition in spite of
charge of the BD black hole provided $\alpha \leq \alpha_{\max }$,
while the solutions have an unstable phase for $\alpha>
\alpha_{\max }$.
\acknowledgments{This work has been supported financially by
Research Institute for Astronomy and Astrophysics of Maragha,
Iran.}

\end{document}